\documentclass[twocolumn,prb,superscriptaddress,showpacs]{revtex4}

\usepackage{graphicx}% Include figure files
\usepackage{dcolumn}% Aligntable columns on decimal point
\usepackage{multirow}
\usepackage{bm}% bold math
\usepackage{times}
\usepackage{color}
\usepackage{amstext}
\usepackage{amsmath}
\usepackage{amsfonts}
\usepackage[bookmarks=true,colorlinks=true,pdfpagemode=UseNone,pdfstartview=Fit,linkcolor=myrefblue,urlcolor=myrefblue,citecolor=myrefblue]{hyperref}
\definecolor{lightblue}{rgb}{0.6,0.9,1}
\definecolor{myrefblue}{rgb}{0.1,0.6,1}
\definecolor{myblue}{rgb}{0,0,0}
\definecolor{nmat}{rgb}{0.7,0.04,0.26}

\begin{document}

\renewcommand{\arraystretch}{1.5}

\title{\mathversion{bold}Spin-density-wave--induced anomalies in the optical conductivity of $A\textrm{Fe}_2\textrm{As}_2$, ($A$=Ca,~Sr,~Ba) single-crystalline iron pnictides\mathversion{normal}}

\author{A.~Charnukha}
\altaffiliation[Currently at: ]{Leibniz Institute for Solid State and Materials Research, IFW, D-01069 Dresden, Germany}
\affiliation{Max-Planck-Institut f\"ur Festk\"orperforschung, Heisenbergstrasse 1, D-70569 Stuttgart, Germany}
\author{D.~Pr\"opper}
\author{T.~I.~Larkin}
\author{D.~L.~Sun}
\author{Z.~W.~Li}
\author{C.~T.~Lin}
\affiliation{Max-Planck-Institut f\"ur Festk\"orperforschung, Heisenbergstrasse 1, D-70569 Stuttgart, Germany}
\author{T.~Wolf}
\affiliation{Institut f\"ur Festk\"orperphysik, Karlsruhe Institute of Technology, D-76021 Karlsruhe, Germany}
\author{B.~Keimer}
\author{A.~V.~Boris}
\affiliation{Max-Planck-Institut f\"ur Festk\"orperforschung, Heisenbergstrasse 1, D-70569 Stuttgart, Germany}

\begin{abstract}
We report the complex dielectric function of high-quality $A\textrm{Fe}_2\textrm{As}_2$, ($A$=Ca,~Sr,~Ba) single crystals with $T_{\mathrm{N}}\approx150\ \textrm{K}$, $200\ \textrm{K}$, and $138\ \textrm{K}$, respectively, determined by broadband spectroscopic ellipsometry at temperatures $10\leq T\leq300\ \textrm{K}$ and wavenumbers from $100\ \textrm{cm}^{-1}$ to $52000\ \textrm{cm}^{-1}$. In $\textrm{CaFe}_2\textrm{As}_2$ we identify the optical spin-density--wave gap $2\Delta_{\mathrm{SDW}}\approx1250\ \textrm{cm}^{-1}$. The $2\Delta_{\mathrm{SDW}}/(k_{\mathrm{B}}T_{\mathrm{N}})$ ratio, characterizing the strength of the electron-electron coupling in the spin-density--wave state, amounts to $\approx12$ in $\textrm{CaFe}_2\textrm{As}_2$, significantly larger than the corresponding values for the $\textrm{SrFe}_2\textrm{As}_2$ and $\textrm{BaFe}_2\textrm{As}_2$ compounds: 8.7 and 5.3, respectively. We further show that, similarly to the Ba-based compound, two characteristic SDW energy gaps can be identified in the infrared-conductivity spectra of both $\textrm{SrFe}_2\textrm{As}_2$ and $\textrm{CaFe}_2\textrm{As}_2$ and investigate their detailed temperature dependence in all three materials. This analysis reveals the existence of an anomaly in $\textrm{CaFe}_2\textrm{As}_2$ at a temperature $T^*\approx80\ \textrm{K}$, well below the N\'eel temperature of this compound, which implies weak coupling between the two SDW subsystems. The coupling between the two subsystems evolves to intermediate in the Sr-based and strong in the Ba-based material. The temperature dependence of the infrared phonons reveals clear anomalies at the corresponding N\'eel temperatures of the investigated compounds. In $\textrm{CaFe}_2\textrm{As}_2$, the phonons exhibit signatures of SDW fluctuations above $T_N$ and some evidence for anomalies at $T^*$. Investigation of all three materials in the visible spectral range reveals a spin-density--wave--induced suppression of two absorption bands systematically enhanced with decreasing atomic number of the intercalant. A dispersion analysis of the data in the entire spectral range clearly shows that $\textrm{CaFe}_2\textrm{As}_2$ is significantly more metallic than the other two compounds. Our results single out $\textrm{CaFe}_2\textrm{As}_2$ in the class of $\textrm{ThCr}_2\textrm{Si}_2$-type iron-based materials by demonstrating the existence of two weakly coupled and extremely metallic electronic subsystems.
\end{abstract}


\maketitle
\section{Introduction}
In the family of iron-based superconductors,~\cite{kamihara} the $\textrm{ThCr}_2\textrm{Si}_2$-type materials are arguably the most widely studied.~\cite{Johnston_Review_2010,RevModPhys.83.1589} This fact owes to the early discovery~\cite{PhysRevB.78.020503,PhysRevLett.101.107006} and synthesis of large high-quality single-crystalline compounds of this class,~\cite{PhysRevLett.101.117004,PhysRevB.79.014506,PhysRevLett.102.117005,PhysRevLett.102.187004} thus allowing the entire arsenal of experimental condensed-matter techniques to be applied to them virtually simultaneously, giving rise to explosive progress in this area of research. The most attention of the community has since been focused on one particular member of the 122 class, $\textrm{BaFe}_2\textrm{As}_2$. In this antiferromagnetic metal superconductivity can be induced by many different means such as electron or hole doping (e.g. by substituting cobalt for iron~\cite{PhysRevB.79.014506} or potassium for barium,~\cite{Rotter_Johrendt_BKFA_phasediagram2008} respectively), isoelectronic substitution (e.g. by substituting ruthenium for iron~\cite{PhysRevB.81.224503,PhysRevB.82.014534} or phosphorus for arsenic~\cite{0953-8984-21-38-382203}) or external pressure.~\cite{Johnston_Review_2010} In all cases there emerges a phase diagram strikingly similar to that of the cuprate high-temperature superconductors,~\cite{BasovChubukov_NatPhys2011} with superconductivity occurring in a dome-shaped region in the immediate vicinity of the antiferromagnetic phase. The development of a spin-density--wave order is accompanied by the opening of an energy gap in the excitation spectrum, which has been confirmed by numerous experimental techniques.~\cite{Johnston_Review_2010} Investigations of the optical conductivity of this and other compounds have been instrumental in obtaining the value of the energy gap and characterising the energetics of the spin-density--wave order.~\cite{PhysRevLett.101.257005,PhysRevB.81.100512,PhysRevB.81.104528,2013arXiv1303.4182O}

Although the growth of $\textrm{CaFe}_2\textrm{As}_2$ (Ref.~\onlinecite{PhysRevB.78.014523}) and its phase diagram with respect to cobalt doping~\cite{PhysRevB.83.094523} were first reported several years ago, the interest in this material started to build up upon the observation of its surprising sensitivity not only to doping and pressure but also to annealing.~\cite{PhysRevB.85.224528} The direct observation of electronic nematicity due to anisotropic impurity scattering potential of substituted cobalt atoms in the underdoped region of the phase diagram of $\textrm{Ca}(\textrm{Fe}_{1-x}\textrm{Co}_x)_2\textrm{As}_2$ (Ref.~\onlinecite{Science327.181,Davis_Nematicity_BFCA_2013}), analogous to $\textrm{Ba}(\textrm{Fe}_{1-x}\textrm{Co}_x)_2\textrm{As}_2$ (Refs.~\onlinecite{Jiun-HawChu08132010,Fisher_BFCA_Nematicity_2012,PhysRevLett.110.207001,PhysRevLett.109.217003}), has brought $\textrm{CaFe}_2\textrm{As}_2$ to the forefront of condensed-matter research. Notwithstanding all this interest and some preliminary measurements~\cite{Nakajima2010S326} (normal-state optical-conductivity data from Ref.~\onlinecite{Nakajima2010S326} were also reproduced in Ref.~\onlinecite{PhysRevB.81.100512}), as well as extensive studies of the charge dynamics in both the parent $(\textrm{Ba},\textrm{Sr},\textrm{Eu})\textrm{Fe}_2\textrm{As}_2$ (Refs.~\onlinecite{PhysRevLett.101.257005,Hu2009545,PhysRevB.79.155103,PhysRevB.81.100512,PhysRevB.84.052501,PhysRevB.86.134503}) and most of their doped/substituted superconducting derivatives,~\cite{PhysRevLett.101.107004,PhysRevB.81.060509,PhysRevB.81.104528,PhysRevB.82.224507,1367-2630-12-7-073036,PhysRevB.82.184527,PhysRevB.82.174509,PhysRevB.82.054518,PhysRevLett.108.147002,PhysRevB.83.100503,PhysRevLett.109.027006,PhysRevB.88.094501,2013arXiv1308.6113N,2013arXiv1308.6133N} the detailed investigation of the charge dynamics in $\textrm{CaFe}_2\textrm{As}_2$ and a systematic comparison with other 122-type materials are still lacking. 

In the present work we fill this gap by reporting broad band ellipsometric study of not only $\textrm{CaFe}_2\textrm{As}_2$ but also its \hbox{Sr-} and Ba-based counterparts and a systematic comparison thereof over a wide spectral range, complemented by a thorough Drude-Lorentz analysis of the interband and itinerant optical response. We find that, similarly to the previous reports on $\textrm{BaFe}_2\textrm{As}_2$ and $\textrm{SrFe}_2\textrm{As}_2$ (Ref.~\onlinecite{PhysRevLett.101.257005,PhysRevLett.108.147002}), two distinct spin-density--wave energy gaps can be identified in the Ca-based compound. In order to address the degree of interplay between these two spin-density--wave subsystems we further carried out a detailed temperature-dependent study of the far-infrared conductivity of all these compounds and discovered the presence of an anomaly at $T*\approx80\ \textrm{K}$ in $\textrm{CaFe}_2\textrm{As}_2$ well below its N\'eel temperature at the frequencies of the smaller spin-density--wave energy gap. This observation clearly indicates that the two electronic subsystems in this material are weakly coupled (as shown for the analogous case of superconductivity in Refs.~\onlinecite{PhysRevB.71.054501,PhysRevB.80.014507,Andreev_two_gaps_FeSe_2011,PhysRevLett.104.087004,Andreev_two_gaps_FeSe_2013}). Further comparison with the \hbox{Sr-} and Ba-based counterparts reveals that this coupling evolves via intermediate in the former to strong in the latter compound, systematically with increasing atomic number of the intercalant. The temperature dependence of the infrared phonons shows a clear anomaly at $T_{\mathrm{N}}$ in all three compounds, whereby the spin-density--wave--induced modifications of the phonon properties set in exactly at $T_{\mathrm{N}}$ in both \hbox{Ba-} and Sr-based materials but at somewhat higher temperatures in $\textrm{CaFe}_2\textrm{As}_2$, suggesting the early development of spin-density--wave fluctuations and their impact on the lattice. We also find a possible indication of $T^*$ in the temperature dependence of the Ca-related phonon intensity $\Delta\varepsilon_0$. 

%%%%%%%%%%%%%%%%%%%%%%%%%%%%%%%%%%%%%%%%%%%%%%%
\begin{figure}[b!]
\includegraphics[width=\columnwidth]{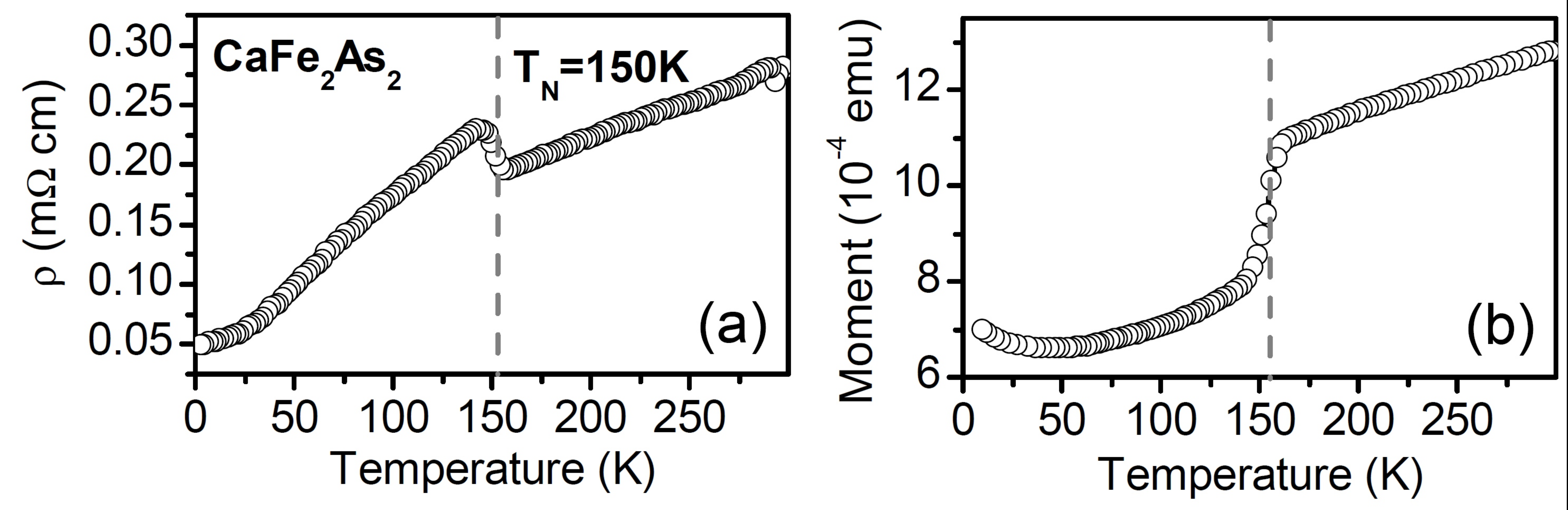}
\caption{\label{fig:characterization} Temperature dependence of the resistivity (a) and magnetic moment (b) of the $\textrm{CaFe}_2\textrm{As}_2$ single crystals used in the present study. Both measurements reveal a clear signature of a spin-density--wave transition at about $150\ \textrm{K}$ and the absence of any secondary phase transitions at lower temperatures.}
\end{figure}
%%%%%%%%%%%%%%%%%%%%%%%%%%%%%%%%%%%%%%%%%%%%%%%%

Investigation of the optical conductivity in the visible spectral range further reveals the existence of a spin-density--wave--induced suppression of two absorption bands, previously reported for $\textrm{SrFe}_2\textrm{As}_2$ in Ref.~\onlinecite{CharnukhaNatCommun2011}, also in the \hbox{Ca-} and Ba-based compounds, with the intensity of the suppression systematically enhanced with decreasing atomic number of the intercalating atom.

Finally, a thorough dispersion analysis reveals a dramatic enhancement of metallicity with decreasing atomic number of the intercalant manifested in the growing total plasma frequency and a systematic evolution of a number of other electronic parameters.

\section{Experimental details}

The parent $(\textrm{Ca,Sr})\textrm{Fe}_2\textrm{As}_2$ single crystals were grown in zirconia crucibles sealed in quartz ampoules under argon atmosphere.~\cite{Lin_BKFA_growth_2010} From dc resistivity and magnetization measurements we obtained the N\'eel temperatures $T_{\mathrm{N}}=150$ and $200\ \textrm{K}$ for $\textrm{CaFe}_2\textrm{As}_2$ (see Fig.~\ref{fig:characterization}) and $\textrm{SrFe}_2\textrm{As}_2$ (Ref.~\onlinecite{Lin_BKFA_growth_2010}), respectively. $\textrm{BaFe}_2\textrm{As}_2$ single crystals were grown from As-rich self-flux in a glassy carbon crucible using prereacted $\textrm{FeAs}_2$ powders mixed with Ba and As.~\cite{PhysRevLett.102.187004} The sample surface was cleaved prior to every optical measurement. The full complex optical conductivity $\sigma(\omega)$ was obtained in the range $100-52000\ \textrm{cm}^{-1}$ using broad~band ellipsometry, as described in Ref.~\onlinecite{boris:027001}. The measurements in the lowest far-infrared spectral range were carried out at the infrared beamline of the ANKA synchrotron light source at Karlsruhe Institute of Technology, Germany.

\section{Results and discussion}
\subsection{\mathversion{bold}Optical conductivity of $A\textrm{Fe}_2\textrm{As}_2$, ($A$=Ca,~Sr,~Ba)\mathversion{normal}}

%%%%%%%%%%%%%%%%%%%%%%%%%%%%%%%%%%%%%%%%%%%%%%%
\begin{figure*}[t!]
\includegraphics[width=\textwidth]{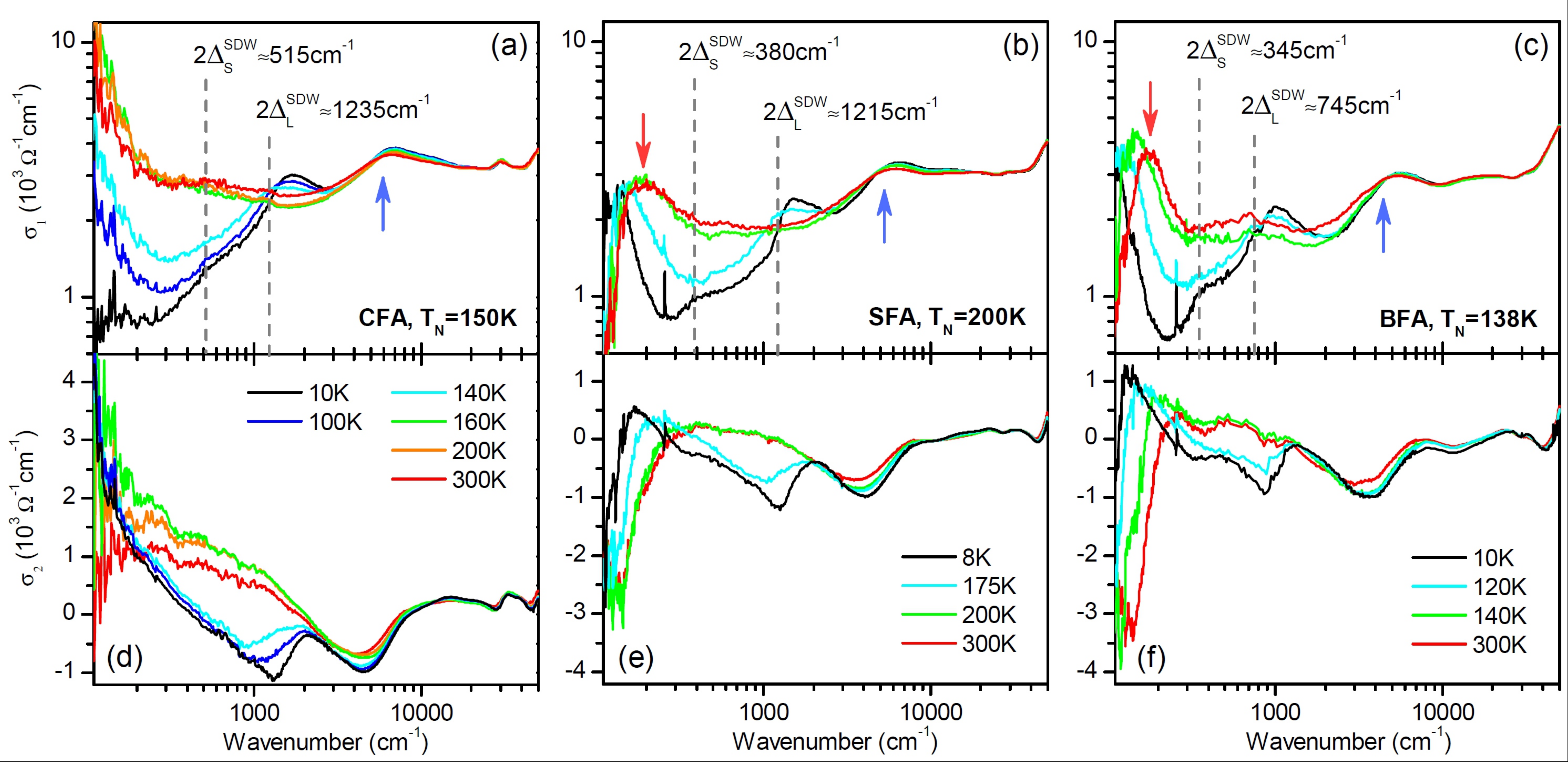}
\caption{\label{fig:rawspectra}(a)--(c)~Real part of the complex optical conductivity of $\textrm{CaFe}_2\textrm{As}_2$~(a), $\textrm{SrFe}_2\textrm{As}_2$~(b), and $\textrm{BaFe}_2\textrm{As}_2$~(c). Characteristic energies of the large and small optical spin-density--wave gaps ($2\Delta_{\mathrm{L,S}}^{\mathrm{SDW}}$, vertical dashed lines) are defined as the intersection points between the lowest and the N\'eel temperatures for the large gap and the deviation point for the small gap (see text). Strongly temperature dependent absorption bands are marked with blue arrows. Low-energy peak feature observed in Sr- and Ba-based compounds but not in $\textrm{CaFe}_2\textrm{As}_2$ is indicated with red arrows. (d)--(f) The corresponding imaginary part of the complex optical conductivity.}
\end{figure*}
%%%%%%%%%%%%%%%%%%%%%%%%%%%%%%%%%%%%%%%%%%%%%%%%

\par The independently obtained real and imaginary parts of the complex optical conductivity $\sigma(\omega)=\sigma_1(\omega)+i\sigma_2(\omega)$ of the $A\textrm{Fe}_2\textrm{As}_2$, ($A$=Ca,~Sr,~Ba) compounds in the entire investigated spectral range at several representative temperatures is shown in Fig.~\ref{fig:rawspectra}. The far-infrared response of $\textrm{CaFe}_2\textrm{As}_2$ [Fig.~\ref{fig:rawspectra}(a) and \ref{fig:rawspectra}(d)] in the normal state is dominated by a strong itinerant response, clearly manifested in the large positive values of both $\sigma_1(\omega)$ and $\sigma_2(\omega)$. Below the spin-density--wave transition temperature of $150\ \textrm{K}$ this itinerant response experiences a dramatic suppression due to the opening of an energy gap in the quasiparticle excitation spectrum, as previously observed in this and other 122-type iron-based materials.~\cite{PhysRevLett.101.257005,PhysRevB.81.100512,PhysRevB.81.104528,2013arXiv1303.4182O} Due to the conservation of the total number of electrons, the missing area under the conductivity curve at low frequences is transferred to higher energies, which results in a formation of a ``hump'' structure.~\cite{JPSJ.12.570,PhysRevB.66.134510} In the most common case when the spin-density--wave energy gap does not cover the entire Fermi surface of a material and, therefore, the optical conductivity remains non-zero at all frequencies, determination of the optical quasiparticle excitation gap $2\Delta$ from the conductivity spectra is difficult but it can be approximated by the energy, at which the low-temperature conductivity spectrum crosses the one at the N\'eel temperature for the first time.~\cite{PhysRevB.66.134510} This energy $2\Delta_{\mathrm{L}}^{\mathrm{SDW}}$ is marked by the right vertical dashed lines in panel~\ref{fig:rawspectra}(a). At lowest frequencies, the optical conductivity of $\textrm{CaFe}_2\textrm{As}_2$ reveals a change in its shape at $10\ \textrm{K}$ as compared to intermediate temperatures below $T_{\mathrm{N}}$, which indicates the presence of another spin-density--wave gap, with a smaller magnitude $2\Delta_{\mathrm{S}}^{\mathrm{SDW}}$ [left vertical dashed line in panel \ref{fig:rawspectra}(a)]. This observation is analogous to the previous report of two different spin-density--wave energy gaps in the \hbox{Sr-}~and Ba-based compound of the same class.~\cite{PhysRevLett.101.257005,Hu2009545,PhysRevLett.108.147002} Our own ellipsometric measurements on the Sr- and Ba-based compounds, shown in Fig.~\ref{fig:rawspectra}(b),\ref{fig:rawspectra}(e) and~\ref{fig:rawspectra}(c),\ref{fig:rawspectra}(f), fully reproduce and confirm previously reported data,~\cite{PhysRevLett.101.257005,PhysRevB.81.100512,PhysRevB.81.104528,2013arXiv1303.4182O,Hu2009545,PhysRevB.84.052501,PhysRevLett.108.147002} provide additional information due to the independently obtained real and imaginary parts of the optical conductivity, and allow for a systematic investigation of the electronic properties of the $\textrm{ThCr}_2\textrm{Si}_2$-type iron-based materials as a function of the intercalating atom.

First of all, we find that, although the N\'eel temperature does not show a systematic dependence on the atomic number of the intercalating atom, the gap ratios, which quantify the strenght of the coupling to a boson mediating a given electronic instability, certainly do display some systematic trends, as shown in Table~\ref{table:sdwgaps}. Here we define the gap ratio of an itinerant antiferromagnetic material with an optical spin-density--wave energy gap $2\Delta$ and a N\'eel temperature $T_{\mathrm{N}}$ as $2\Delta/k_{\mathrm{B}}T_{\mathrm{N}}$, where $k_{\mathrm{B}}$ is the Boltzmann constant. The values of the large and small optical spin-density--wave gaps are extracted from the experimentally obtained optical conductivity as described above and indicated in Fig.~\ref{fig:rawspectra} with vertical dashed lines.

%%%%%%%%%%%%%%%%%%%%%%%%%%%%%%%%%%%%%%%%%%%%%%%
\begin{figure*}[tb]
\includegraphics[width=\textwidth]{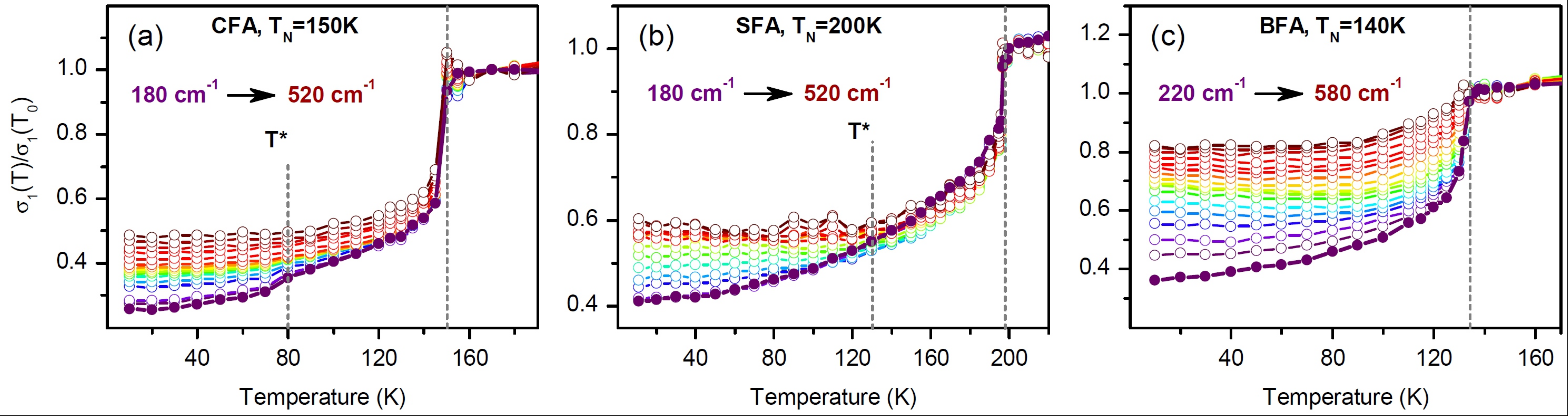}
\caption{\label{fig:sdwgaps}(a)--(c)~Detailed temperature dependence of the real part of the complex optical conductivity of $\textrm{CaFe}_2\textrm{As}_2$~(a), $\textrm{SrFe}_2\textrm{As}_2$~(b), and $\textrm{BaFe}_2\textrm{As}_2$~(c) on a fine grid of wavenumbers in the far-infrared spectral range, normalized to the value of the optical conductivity at $T=T_{\mathrm{0}}$ close to $T_{\mathrm{N}}$ ($T_{\mathrm{0}}=170\ \textrm{K},\ 200\ \textrm{K},\ \textrm{and}\ 140\ \textrm{K}$ for the Ca-, Sr-, and Ba-based compounds, respectively). The grid of wavenumbers runs from $180\ \textrm{cm}^{-1}$ (blue colors) to $520\ \textrm{cm}^{-1}$ (red colors) in steps of $20\ \textrm{cm}^{-1}$ with the exception of the spectral window of the $260\ \textrm{cm}^{-1}$ phonon. Spin-density--wave--related features in the temperature dependence of the optical conductivity (vertical dashed lines), for interpretation see text. The lowest-frequency temperature dependence is plotted as filled circles unlike the rest (open circles) to emphasize the change in the temperature dependence of the optical conductivity.}
\end{figure*}
%%%%%%%%%%%%%%%%%%%%%%%%%%%%%%%%%%%%%%%%%%%%%%%%

Table~\ref{table:sdwgaps} shows that, albeit the Ca-based compound has a N\'eel temperature significantly lower than that of the Sr-based counterpart and comparable to that of $\textrm{BaFe}_2\textrm{As}_2$, its both gap ratios are significantly larger than those of the other two materials, which implies a much stronger electronic instability in this compound. In addition, the larger gap ratio $2\Delta_{\mathrm{L}}^{\mathrm{SDW}}/k_{\mathrm{B}}T_{\mathrm{N}}$ gradually decreases with the increasing atomic number of the intercalant.

\begin{table}[b!]
\caption{\label{table:sdwgaps}Large and small optical spin-density--wave gap $2\Delta_{\mathrm{L,S}}^{\mathrm{SDW}}$ as inferred from the optical conductivity spectra in Fig.~\ref{fig:rawspectra} and the corresponding gap ratios.}
\begin{tabular*}{\columnwidth}{@{\extracolsep{\fill}} l c c c}
\hline
&$\textrm{CaFe}_2\textrm{As}_2$&$\textrm{SrFe}_2\textrm{As}_2$&$\textrm{BaFe}_2\textrm{As}_2$\\
\hline
$2\Delta_{\mathrm{L}}^{\mathrm{SDW}},\ \textrm{cm}^{-1}$&1235&1215&745\\
$2\Delta_{\mathrm{S}}^{\mathrm{SDW}},\ \textrm{cm}^{-1}$&515&380&345\\
$k_{\mathrm{B}}T_{\mathrm{N}},\ \textrm{cm}^{-1}$&105&140&97\\
\hline
$2\Delta_{\mathrm{L}}^{\mathrm{SDW}}/k_{\mathrm{B}}T_{\mathrm{N}}$&11.8&8.7&7.7\\
$2\Delta_{\mathrm{S}}^{\mathrm{SDW}}/k_{\mathrm{B}}T_{\mathrm{N}}$&4.9&2.7&3.6\\
$2\Delta_{\mathrm{L}}^{\mathrm{SDW}}/2\Delta_{\mathrm{S}}^{\mathrm{SDW}}$&2.4&3.2&2.14\\
\hline
\end{tabular*}
\end{table}

The far-infrared optical conductivity of $\textrm{SrFe}_2\textrm{As}_2$ and $\textrm{BaFe}_2\textrm{As}_2$ at lowest frequencies is dominated by a peak feature [red arrows in Figs.~\ref{fig:rawspectra}(b),\ref{fig:rawspectra}(c)] reminiscent of the collective excitation previously observed in the nearly optimally doped high-temperature cuprate superconductor $\textrm{YBa}_2\textrm{Cu}_3\textrm{O}_{6+x}$ pristine~\cite{PhysRevB.69.052502} and after irradiation with helium ions.~\cite{PhysRevB.49.12165} A similar albeit weaker feature has also been identified in the far-infrared optical response of superconducting $\textrm{Ba}(\textrm{Fe}_{0.92}\textrm{Co}_{0.08})_2\textrm{As}_2$ in Ref.~\onlinecite{PhysRevB.82.100506}. One of the possible origins of this feature is charge-carrier localization, as observed in disordered doped superconductors.~\cite{PhysRevB.33.7329,PhysRevB.25.3519,PhysRevB.41.5152}

At higher energies the optical conductivity of all three compounds shows a strongly temperature-dependent absorption band centered at about 5000--6000~$\textrm{cm}^{-1}$, as well as at UV energies in the case of $\textrm{BaFe}_2\textrm{As}_2$, as shown in Figs.~\ref{fig:rawspectra}(a)-\ref{fig:rawspectra}(c) with blue arrows.

\subsection{Temperature dependence of the SDW-induced suppression}

The existence of two seperate spin-density--wave energy gaps raises a question as to the degree of interaction between the corresponding electronic subsystems. This question becomes particularly difficult to address in the presence of as many Fermi-surface sheets as is the case for the iron-based materials, which often commonly possess five separate Fermi-surface sheets in the first Brillouin zone. This complexity arises mainly due to the fact that the optical conductivity represents an intertwined response of all electronic subsystems at once. To shed some light onto this issue, in addition to our broad band ellipsometric measurements at several representative temperatures, we have carried out detailed measurements in the far-infrared spectral range (i.e. in the range where the real part of the optical conductivity experiences a drastic suppression, as shown in Fig.~\ref{fig:rawspectra}) with a very fine temperature grid for all three materials in question. The results are compiled in Fig.~\ref{fig:sdwgaps}, which shows the temperature dependence of the optical conductivity normalized to its value at $T=T_{\mathrm{0}}$ close to $T_{\mathrm{N}}$ ($T_{\mathrm{0}}=170\ \textrm{K},\ 200\ \textrm{K},\ \textrm{and}\ 140\ \textrm{K}$ for the Ca-, Sr-, and Ba-based compounds, respectively) at an equidistant set of wavenumbers from $180\ \textrm{cm}^{-1}$ (blue colors) to $520\ \textrm{cm}^{-1}$ (red colors) in steps of $20\ \textrm{cm}^{-1}$ with the exception of the spectral window of the $260\ \textrm{cm}^{-1}$ phonon.

The temperature dependence of the normalized conductivity shows a very pronounced drop for all compounds at their respective N\'eel temperatures (right vertical dashed lines) due to the onset of spin-density--wave order, with a clear mean-field order-parameter--like temperature dependence at large wavenumbers (red colors). This temperature dependence changes, however, as one moves toward progressively smaller wavenumbers and in the case of $\textrm{CaFe}_2\textrm{As}_2$ reveals a clear second suppresssion feature at a temperature $T^*$ of about $80\ \textrm{K}$. This second feature becomes washed out in the Sr-based compound, although the temperature dependence of $\sigma_1$ at smallest wavenumbers remains markedly different from that at large wavenumbers and some sort of an analogue of $T^*$ can be sketched also in this case, although there is no clear second suppression feature present in any of the separate temperature dependences themselves. Finally, in the case of $\textrm{BaFe}_2\textrm{As}_2$, the change in the temperature dependence of $\sigma_1$ between large and small wavenumbers becomes hardly noticeable, with minute changes below $\approx60\ \textrm{K}$.

The presence of a second suppression feature in the temperature dependence of the real part of the optical conductivity at small enough wavenumbers in $\textrm{CaFe}_2\textrm{As}_2$ indicates, that the electronic subsystem that develops the smaller spin-density--wave energy gap $2\Delta_{\mathrm{S}}^{\mathrm{SDW}}$ preserves some knowledge about its own N\'eel temperature that it {\it would have} if this subsystem were completely independent from the other(s). In any real material all electronic subsystems are coupled, even if weakly, so that all spin-density--wave energy gaps open at the same $T_{\mathrm{N}}$. However, in case of weak intersubsystem coupling, those with smaller gaps exhibit an anomaly below the real N\'eel temperature, as is the case in $\textrm{CaFe}_2\textrm{As}_2$. Such an effect has been predicted theoretically for the analogous case of multiband superconductivity~\cite{PhysRevB.71.054501,PhysRevB.80.014507} and discovered experimentally in $\textrm{FeSe}_{1-x}$ in Refs.~\onlinecite{Andreev_two_gaps_FeSe_2011,PhysRevLett.104.087004,Andreev_two_gaps_FeSe_2013}. This argument thus suggests that in the Ca-based material the two electronic subsystems developing the large and the small spin-density--wave energy gap are weakly coupled. Naturally, as the coupling between such electronic subsystems increases, the temperature dependence of the small gap would gradually approach that of the large gap via an intermediate state when it already does not show any anomaly but still has not matched the temperature dependence of the large gap. Figure~\ref{fig:sdwgaps}(b) provides evidence for such a behavior in the Sr-based iron pnictide. Finally, the temperature dependence of the normalized optical conductivity in $\textrm{BaFe}_2\textrm{As}_2$ is almost the same at all wavenumbers, both in the region of the large spin-density--wave gap and in that of the small one, indicating strongly coupled electronic subsystems in this compound. Thus, by monitoring the detailed temperature dependence of the optical conductivity in the far-infrared spectral range on a fine grid of wavenumbers, a gradual transition could be observed from weak coupling between the electronic subsystems developing the two different spin-density--wave energy gaps in $\textrm{CaFe}_2\textrm{As}_2$, via intermediate coupling in $\textrm{SrFe}_2\textrm{As}_2$, to strong coupling between them in $\textrm{BaFe}_2\textrm{As}_2$, systematically with increasing atomic number of the intercalant.

%%%%%%%%%%%%%%%%%%%%%%%%%%%%%%%%%%%%%%%%%%%%%%%
\begin{figure}[tb]
\includegraphics[width=\columnwidth]{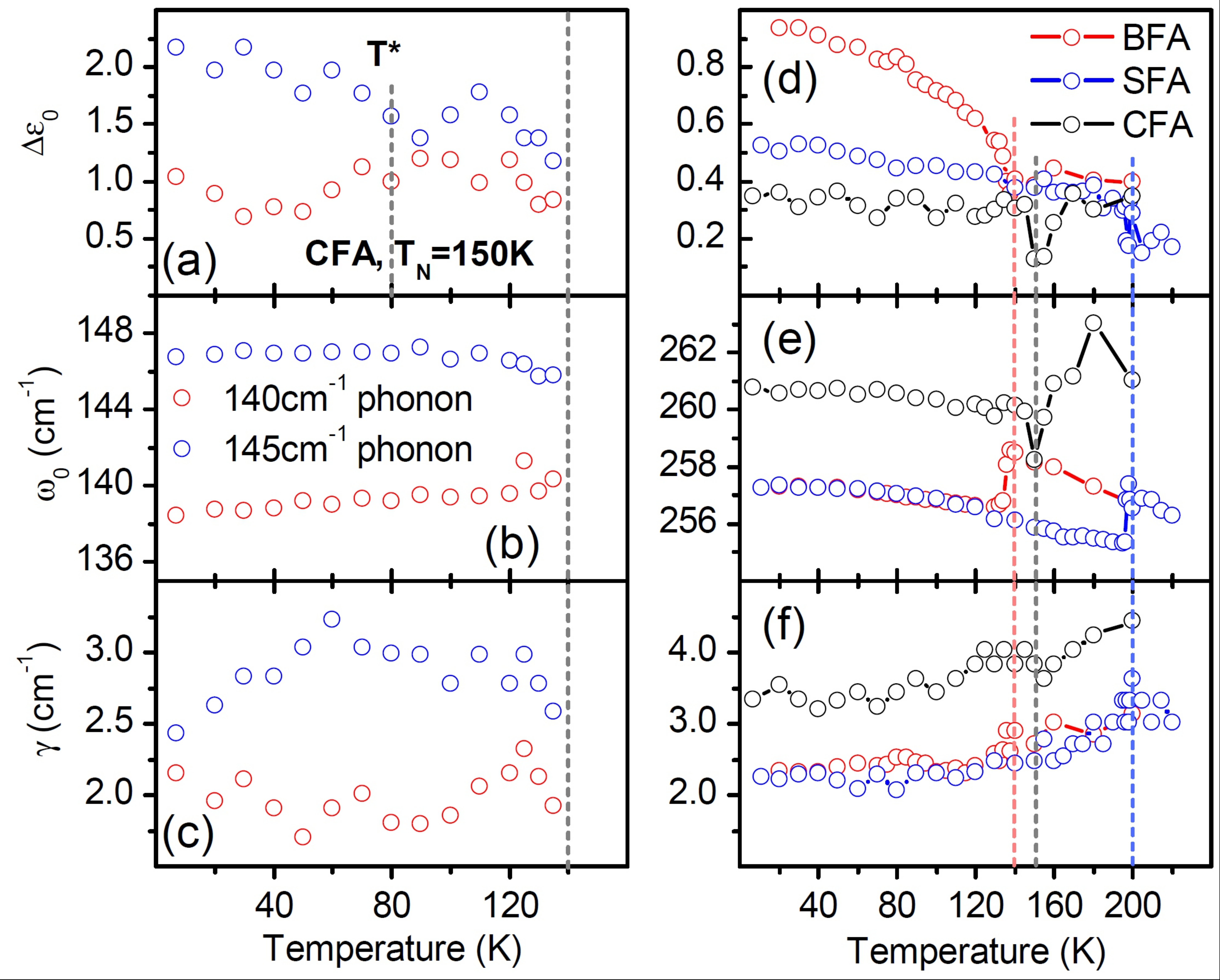}
\caption{\label{fig:phonons}(a)--(c)~Temperature dependence of the strength $\Delta\varepsilon$, position $\omega_0$, and the width $\gamma$ of the Ca-related phonon below the N\'eel transition temperature of $\textrm{CaFe}_2\textrm{As}_2$, split due to the lowering of the crystallographic symmetry from tetragonal to orthorhombic at the concomitant structural transition (right vertical dashed line). The temperature $T^*$ of the small optical spin-density--wave gap $2\Delta_{\mathrm{S}}^{\mathrm{SDW}}$ [left vertical dashed line in (a)] as inferred from Fig.~\ref{fig:sdwgaps}(a). (d)--(f) Temperature dependence of the strength $\Delta\varepsilon$, position $\omega_0$, and the width $\gamma$ of the phonon due to the vibrations of Fe and As ions. Vertical dashed lines indicate the N\'eel transition temperatures of the Ca-, Sr-, and Ba-based compounds (grey, light blue, pink, respectively).}
\end{figure}
%%%%%%%%%%%%%%%%%%%%%%%%%%%%%%%%%%%%%%%%%%%%%%%%

\subsection{Temperature dependence of the far-infrared phonons}

We now use the same detailed ellipsometric measurements of the real and imaginary parts of the optical conductivity to analyze and compare between the three different compounds the temperature dependence of the strength, position, and the width of the far-infrared optical phonons. As can be seen in Fig.~\ref{fig:rawspectra}, the low-energy phonon due to the vibrations of the intercalating ions can be seen at $\approx140\ \textrm{cm}^{-1}$ in the Ca-based compound but not in the Sr- and Ba-based counterparts due to the much smaller mass of the former, which pushes this phonon into the accessible experimental spectral range. The phonon arising from vibrations of Fe and As is seen at an approximately the same position of $260\ \textrm{cm}^{-1}$ in all three materials. The detailed temperature dependence of the Lorentz parameters of the Ca-related phonon at temperatures below $T_{\mathrm{N}}$, where it can be clearly resolved, as well as the comparision of the Lorentz parameters of the $260\ \textrm{cm}^{-1}$-phonon in all three compounds is shown in Fig.~\ref{fig:phonons}.

%%%%%%%%%%%%%%%%%%%%%%%%%%%%%%%%%%%%%%%%%%%%%%%
\begin{figure*}[t!]
\includegraphics[width=\textwidth]{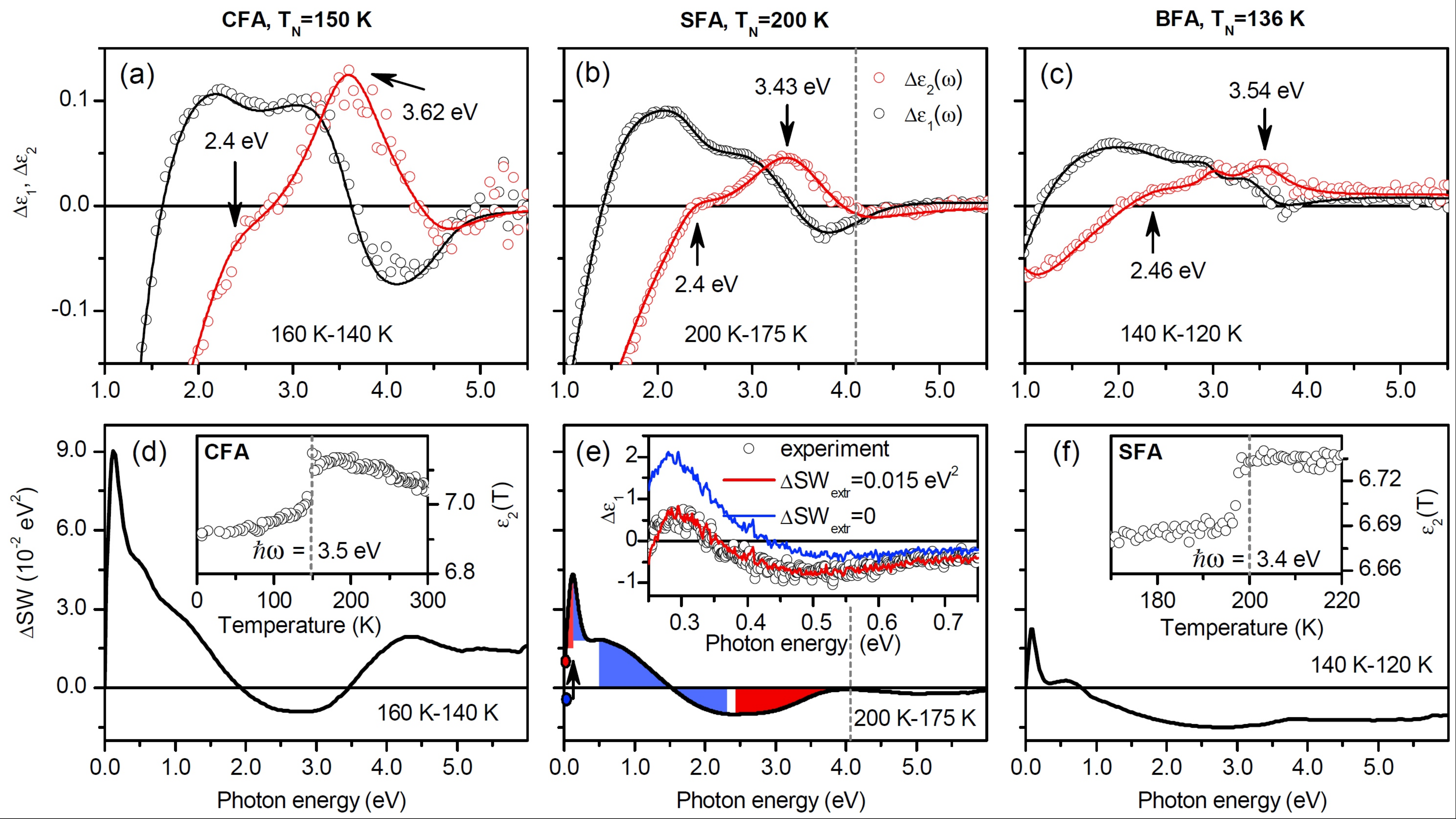}
\caption{\label{fig:anomalyvis}(a)--(c)~Difference real and imaginary parts of the dielectric function $\Delta\varepsilon_1,\ \Delta\varepsilon_2$ in the visible spectral range between the temperatures above and below the N\'eel transition temperature, as specified in the panels. The arrows indicate the two spin-density--wave--supressed absorption bands. (d)--(f) Spectral-weight redistribution between the same temperatures and in the same compounds as in (a)--(c) as a function of photon energy. Blue and red areas in (e) indicate the regions of spectral-weight gain and loss, respectively, in the magnetic versus the normal state. [Inset in (e)], Real part of the dielectric function and the Kramers-Kronig transformations of the real part of the optical conductivity of $\textrm{SrFe}_2\textrm{As}_2$ with different amount of spectral weight contained in the extrapolation region at low frequencies (solid lines, colors match filled circles). [Insets in (d),(f)] Temperature scan of $\varepsilon_2$ in $\textrm{CaFe}_2\textrm{As}_2$ at $3.5\ \textrm{eV}$ and in $\textrm{SrFe}_2\textrm{As}_2$ at $3.4\ \textrm{eV}$, respectively. Panel (e), the inset therein, as well as in (f) adapted by permission from Macmillan Publishers Ltd: Nature Communications Ref.~\onlinecite{CharnukhaNatCommun2011}, copyright (2011).}
\end{figure*}
%%%%%%%%%%%%%%%%%%%%%%%%%%%%%%%%%%%%%%%%%%%%%%%%

In the $\textrm{CaFe}_2\textrm{As}_2$ compound, the splitting of the Ca-related phonon at the magnetostructural transition can be clearly resolved, similarly to the previously reported splitting of the Ba-related phonon in the sister compound,~\cite{PhysRevB.84.052501} and amounts to about $8\ \textrm{cm}^{-1}$ [Fig.~\ref{fig:phonons}(b)] . Within the noise floor the temperature dependence of the width of these two phonons, plotted in Fig.~\ref{fig:phonons}(c), does not appear to display any anomalies. That of the phonon strength $\Delta\varepsilon$, on the other hand, seems to slighly change at the temperature $T^*$ [left vertical dashed line in (a)] inferred from Fig.~\ref{fig:sdwgaps}(a), albeit this change is quite close to the limit of the fit uncertainty.

The temperature dependence of all Lorentz parameters of the Fe-As phonon in all three compounds, Fig.~\ref{fig:phonons}(d)--(f), shows noticeable anomalies at the respective N\'eel temperatures (grey, light blue, and pink vertical dashed lines for \hbox{Ca-}, \hbox{Sr-}, and Ba-based compounds, respectively). In $\textrm{SrFe}_2\textrm{As}_2$ and $\textrm{BaFe}_2\textrm{As}_2$ the phonon intensity $\Delta\varepsilon_0$ and position $\omega_0$ change abruptly at the magnetostructural transition temperature, as shown in Fig.~\ref{fig:phonons}(d),~\ref{fig:phonons}(e), whereas in $\textrm{CaFe}_2\textrm{As}_2$, quite surprisingly, the spin-density--wave--induced changes seem to set in already at somewhat higher temperatures [Fig.~\ref{fig:phonons}(e)]. This observation indicates the existence of incipient critical lattice strain at temperatures higher than the magnetostructural transition temperature in this compound, which may be related to the electronic nematicity recently discovered in a doped compound of this class.~\cite{Davis_Nematicity_BFCA_2013} In addition, while the width and the position of the Fe-As phonon in both Sr- and Ba-doped compounds [Fig.~\ref{fig:phonons}(e),~\ref{fig:phonons}(f)] display essentially identical magnitude and behavior, those of the Ca-doped material are significantly larger. While the somewhat broader phonon feature could, in principle, be traced back to the quality of the sample surface or the sample itself, the harder Fe-As phonon of Ca-based compound compared to its Sr- and Ba-based counterparts must be intrinsic. Such a hardening most likely results from the shorter Ca-As and Fe-As bond lengths compared to those in the \hbox{Sr-} and Ba-based compounds and has also been observed in inelastic-neutron-scattering measurements on $\textrm{CaFe}_2\textrm{As}_2$, $\textrm{BaFe}_2\textrm{As}_2$ and predicted by ab-initio calculations~\cite{PhysRevB.79.144516}.

%%%%%%%%%%%%%%%%%%%%%%%%%%%%%%%%%%%%%%%%%%%%%%%
\begin{figure*}[t!]
\includegraphics[width=\textwidth]{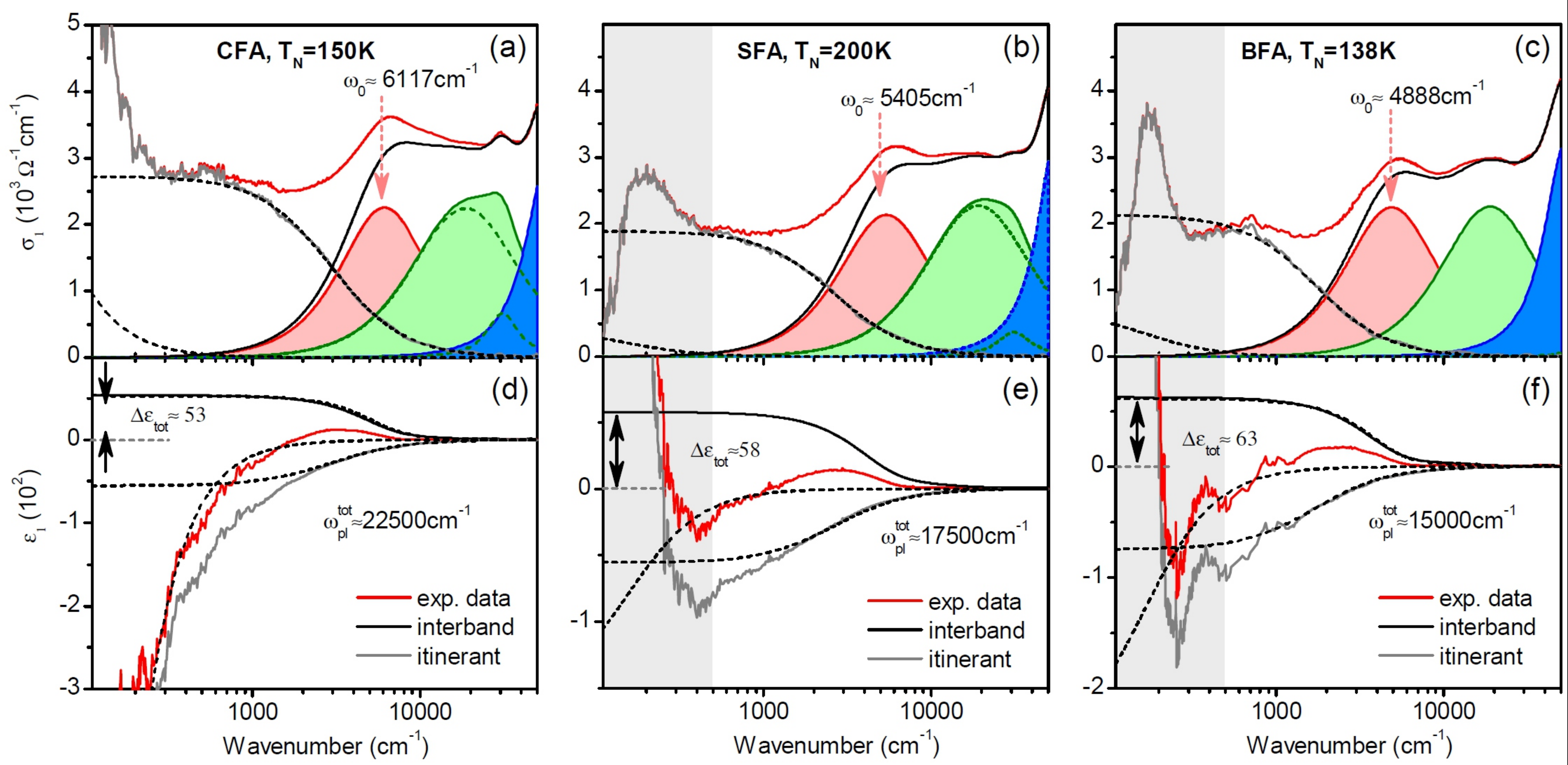}
\caption{\label{fig:dispan} Real part of the optical conductivity~(a)--(c) and dielectric function~(d)--(f) of the \hbox{Ca-}, \hbox{Sr-}, and Ba-based compounds at $300\ \textrm{K}$ (red line) and the corresponding interband [red, green, and blue areas in (a)--(c)] and itinerant (grey line) contributions obtained in a Drude-Lorentz dispersion analysis as described in the text. The partial Drude contributions are shown as black dashed lines. The lowest clearly identifiable absorption band is indicated with a pink arrow. Major partial Lorentz contributions to the absorption band in the visible (green area) are plotted as green dashed lines in (a)--(c). The total contribution of all interband transitions (black line). The grey-shaded area in~(b),~(c) and~(e),~(f) indicates the spectral range excluded from the Drude-Lorentz fit due to the presence of a low-energy collective excitation.}
\end{figure*}
%%%%%%%%%%%%%%%%%%%%%%%%%%%%%%%%%%%%%%%%%%%%%%%%

\subsection{Spin-density--wave--induced anomaly in the visible}

Having investigated the spin-density--wave--induced anomalies in the infrared spectral range, let us turn to the charge dynamics in the 122 compounds at visible frequencies.~\footnote{To facilitate the comparison with previous work, in this section we use electron-volt instead of wavenumber as the unit of photon energy. The conversion factor between the two is $8066\ \textrm{cm}^{-1}/\textrm{eV}$.} Recently, a criticality-induced suppression of an absorption band at energies, much higher than the corresponding energy gaps, has been reported in both the optimally doped superconducting $\textrm{Ba}_{0.68}\textrm{K}_{0.32}\textrm{Fe}_2\textrm{As}_2$ and the antiferromagnetic $\textrm{SrFe}_2\textrm{As}_2$ (Ref.~\onlinecite{CharnukhaNatCommun2011}). The analysis of the spectral-weight transfer across the spin-density--wave transition carried out in the same work, revealed that high energies up to $4\ \textrm{eV}$ are involved in the spectral-weight redistribution and that the $f$-sum rule is fully satisfied above the suppressed absorption band in the visible spectral range. It appears to be of interest whether in other parent compounds of the same 122-type the spin-density--wave energetics is similar or rather shows a systematic variation with the atomic number of the intercalant within the class. To this end we have carried out accurate ellipsometric measurements in the visible spectral range to mach the data quality of Ref.~\onlinecite{CharnukhaNatCommun2011} and compare the intensity and location of the spin-density--wave--supressed absorption bands, as well as the spectral-weight redistribution in the entire investigated spectral range, in all three parent compounds.

The results are presented in Fig.~\ref{fig:anomalyvis}. Panels~\ref{fig:anomalyvis}(a)--(c) show the difference real and imaginary parts of the dielectric function $\varepsilon(\omega)=1+4\pi i\sigma(\omega)/\omega$ in the visible spectral range. The difference is taken between the temperatures above and below the material's N\'eel transition temperature: $160\ \textrm{K}$ and $140\ \textrm{K}$, $200\ \textrm{K}$ and $175\ \textrm{K}$, and $140\ \textrm{K}$ and $120\ \textrm{K}$ for $\textrm{CaFe}_2\textrm{As}_2$, $\textrm{SrFe}_2\textrm{As}_2$, and $\textrm{BaFe}_2\textrm{As}_2$, respectively. These spectra reveal a clear suppression of absorption at optical frequencies. The magnetostructurally induced character of this suppression is confirmed by drastic drop in the temperature dependence of the imaginary part of the dielectric function $\varepsilon_2$ at the N\'eel temperature, shown in the inset of Figs.~\ref{fig:anomalyvis}(d),~\ref{fig:anomalyvis}(f) for the Ca- and Sr-based compounds, respectively. The suppressed optical response of all three iron pnictides shown in Figs.~\ref{fig:anomalyvis}(a)--(c) reveals the contribution of two absorption bands and the systematic evolution of their suppression intensities with the atomic number: the suppression of both bands reduces upon going from lighter to heavier intercalating atoms.

\renewcommand{\arraystretch}{1.2}

\begin{table*}[t!]
\caption{\label{table:drudelorentz}Parameters of the Drude and Lorentzian terms in~(\ref{eq:drudelorentz}) obtained from the dispersion analysis of the optical conductivity of $\textrm{CaFe}_2\textrm{As}_2$, $\textrm{SrFe}_2\textrm{As}_2$, and $\textrm{BaFe}_2\textrm{As}_2$ at room and N\'eel temperature (shown for room temperature in Fig.~\ref{fig:dispan}).}
\begin{tabular*}{\textwidth}{@{\extracolsep{\fill}} c | c c c | c c c | c c c}
\hline
&\multicolumn{3}{c|}{$\textrm{CaFe}_2\textrm{As}_2$}&\multicolumn{3}{c|}{$\textrm{SrFe}_2\textrm{As}_2$}&\multicolumn{3}{c}{$\textrm{BaFe}_2\textrm{As}_2$}\\
\hline
j&$\Delta\varepsilon_j$&$\omega_{0j}\ (\omega_{\mathrm{pl},j}),\ \textrm{cm}^{-1}$&$\Gamma_j\ (\gamma_j),\ \textrm{cm}^{-1}$&$\Delta\varepsilon_j$&$\omega_{0j}\ (\omega_{\mathrm{pl},j}),\ \textrm{cm}^{-1}$&$\Gamma_j\ (\gamma_j),\ \textrm{cm}^{-1}$&$\Delta\varepsilon_j$&$\omega_{0j}\ (\omega_{\mathrm{pl},j}),\ \textrm{cm}^{-1}$&$\Gamma_j\ (\gamma_j),\ \textrm{cm}^{-1}$\\
\hline
\multicolumn{10}{c}{$300\ \textrm{K}$}\\
\hline
Drude$_1$&---&4400&40&---&1960&156&---&2622&162\\
Drude$_2$&---&22000&2960&---&17340&2650&---&14833&1717\\
Drude$_{\mathrm{tot}}$&---&{\bf 22500}&---&---&{\bf 17450}&---&---&{\bf 15060}&---\\
Drude$_{\mathrm{KK}}$&---&{\bf 22000}&---&---&{\bf 17895}&---&---&{\bf 15300}&---\\
&&&&&&&&&\\
L$_1$&{\bf 35}&{\bf 6121}&9683&{\bf 40}&{\bf 5407}&9167&{\bf 46}&{\bf 4888}&8200\\
L$_2$&14.25&18592&36627&13.8&19138&37238&12.7&18930&33539\\
L$_3$&0.68&31035&16374&0.397&31325&17580&0.27&31483&20992\\
L$_4$&2.35&58810&44568&2.47&58087&39548&2.57&53464&38045\\
L$_5$&1.35&113407&3991&1.04&165782&3102&1&80467&3011\\
\hline
\multicolumn{10}{c}{$T_{\mathrm{N}}$}\\
\hline
Drude$_1$&---&5900&40&---&2150&144&---&2300&600\\
Drude$_2$&---&20475&2724&---&17315&2785&---&13212&1310\\
Drude$_{\mathrm{tot}}$&---&21308&---&---&17448&---&---&13410&---\\
Drude$_{\mathrm{KK}}$&---&20870&---&---&17915&---&---&13335&---\\
&&&&&&&&&\\
L$_1$&33.7&6353&9503&37.3&5614&8928&45&4986&7956\\
L$_2$&14.8&18096&35971&13.98&19138&37238&12.5&19460&34555\\
L$_3$&0.7&30703&13062&0.39&31322&16222&0.2&32164&14191\\
L$_4$&2.36&58461&44615&2.44&58087&38761&2.2&52906&33500\\
L$_5$&1.64&273090&8172&0.92&138292&2947&1.12&70949&22284\\
\hline
\end{tabular*}
\end{table*}

We now move to the analysis of the spectral-weight redistribution across the spin-density--wave transition presented in Figs.~\ref{fig:anomalyvis}(d),~~\ref{fig:anomalyvis}(e), and~\ref{fig:anomalyvis}(f) for the case of \hbox{Ca-}, \hbox{Sr-}, and Ba-based material, respectively. The energy-dependent spectral weight is defined as the integral of the real part of the optical conductivity $\sigma_1(\omega)$ up to a certain cut-off frequency $\Omega$:\[SW(\Omega)=\int_0^{\Omega}\sigma_1(\omega)d\omega.\] Several common features can be identified in the energetics of the spin-density--wave transition: the spectral weight from below the optical spin-density--wave gap $2\Delta^{\mathrm{SDW}}$ [lowest-energy red area in Fig.~\ref{fig:anomalyvis}(e) and the corresponding areas in Figs.~\ref{fig:anomalyvis}(d),(f)] is redistributed to energies directly above the optical gap [subsequent blue area in Fig.~\ref{fig:anomalyvis}(e) and the corresponding areas in Figs.~\ref{fig:anomalyvis}(d),(f)]. However, the spectral weight gained directly above the optical gap does not balance that lost within the gap, which implies that higher-energy processes beyond the characteristic magnetic energy scales proposed for these materials are affected by the spin-density--wave transition, including the spin-density--wave--supressed band in the visible spectral range [higher-energy blue and red areas in Fig.~\ref{fig:anomalyvis}(e) and the corresponding areas in Figs.~\ref{fig:anomalyvis}(d),(f)]. Such a high-energy modification of the spectral weight might originate in a redistribution of the electronic population between several bands at the spin-density--wave transition, as suggested for $\textrm{SrFe}_2\textrm{As}_2$ in Ref.~\onlinecite{CharnukhaNatCommun2011}.

These commonalities notwithstanding, it is clear from Figs.~\ref{fig:anomalyvis}(d)--(f) that the overall intensity of the spectral-weight redistribution at the spin-density--wave transition decreases systematically with increasing atomic number of the intercalant even though the corresponding N\'eel temperatures do not reflect this behavior ($T_{\mathrm{N}}$ is almost the same in the~\hbox{Ca-} and Ba-based compound, whereas the amplitude of the spectral-weight redistribution is more than three times larger).

\subsection{Itinerant properties}

Systematic trends similar to those found in the overall energetics of the 122 parent compounds investigated in this work can also be identified in the itinerant-charge-carrier response. However, whereas the former could be revealed already in the raw ellipsometric data, the extration of the properties of the latter requires careful elimination of the interband contribution to the optical conductivity by means of a dispersion analysis (e.g. as carried out in Ref.~\onlinecite{PhysRevB.84.174511}), which allows one to isolate the inherent itinerant response. In the present work we have analyzed the free-charge-carrier and interband contributions in the standard Drude-Lorentz model. 

In this method, the optical conductivity $\sigma(\omega)=\sigma_1(\omega)+i\sigma_2(\omega)$ or, equivalently, the full complex dielectric function $\varepsilon(\omega)=1+4\pi i\sigma(\omega)/\omega$ (in CGS units) is fitted with an expression of the following form:
\begin{equation}%
\varepsilon(\omega)=-\sum\frac{\omega_{\mathrm{pl},j}^2}{\omega^2+i\gamma_j\omega}+\sum\frac{\Delta\varepsilon_j\omega_{0j}^2}{(\omega_{0j}^2-\omega^2)-i\Gamma_j\omega}\label{eq:drudelorentz},
\end{equation}where the first sum runs over all Drude and the second one over all Lorentz terms. $\omega_{\mathrm{pl},j}$ and $\gamma_j$ are the plasma frequency and the quasiparticle scattering rate of the partial itinerant response and $\Delta\varepsilon_j,\ \omega_{0j},\ \Gamma_j$ are the static permittivity contribution, the center frequency and the width of the Lorentzian oscillators used to model the interband transitions, respectively. The function in Eq.~\ref{eq:drudelorentz} is fitted simultaneously to the real and imaginary parts of the dielectric function.

The oscillator parameters obtained in such a fit for all three 122 parent compounds are summarized in Table~\ref{table:drudelorentz}. The contribution of the most important interband transitions to the real part of the optical conductivity is broken down in Figs.~\ref{fig:dispan}(a)--\ref{fig:dispan}(c) (red, green, and blue shaded areas) for the \hbox{Ca-}, \hbox{Sr-}, and Ba-based compounds, respectively, and their integral contribution to the real part of both the optical conductivity [black lines in Figs.~\ref{fig:dispan}(a)--\ref{fig:dispan}(c)] and the dielectric function [black lines in Figs.~\ref{fig:dispan}(d)--\ref{fig:dispan}(f)], as well as the extracted itinerant response (grey lines in all panels), is also shown. The overall structure of the interband transitions is very similar in all three compounds. The lowest clearly identifiable interband transition, previously found to be strongly renormalized with respect to the prediction of band-structure calculations and to give an unusually large contribution to the total zero-frequency interband permittivity $\Delta\varepsilon_{\mathrm{tot}}$ of the order of 60 in the optimally doped $\textrm{Ba}_{1-x}\textrm{K}_x\textrm{Fe}_2\textrm{As}_2$ (see Ref.~\onlinecite{PhysRevB.84.174511}), shows a systematic softening with increasing atomic number of the intercalating atom, with a concomitant increase in $\Delta\varepsilon$. The evolution of the total zero-frequency interband permittivity $\Delta\varepsilon_{\mathrm{tot}}$ is relatively small and is dominated by that of the lowest interband transition, as can be inferred from the limiting zero-frequency behavior of the total interband contribution to the real part of the dielectric function shown with black lines in Figs.~\ref{fig:dispan}(d)--\ref{fig:dispan}(f) and from Table~\ref{table:drudelorentz}.

The extracted itinerant response (grey lines in Fig.~\ref{fig:dispan}) can be broken down into a narrow and broad component (see Table~\ref{table:drudelorentz}) for all three compounds, in agreement with Ref.~\onlinecite{PhysRevB.81.100512}, which previously identified this phenomenon in essentially all known parent iron pnictides of the $\textrm{ThCr}_2\textrm{Si}_2$ type. In our Drude-Lorentz fit we excluded the lowest far-infrared spectral range strongly affected by the collective excitation in the \hbox{Sr-} and Ba-based compounds [gray shaded areas in Figs.~\ref{fig:dispan}(b),~\ref{fig:dispan}(c),~\ref{fig:dispan}(e), and~\ref{fig:dispan}(f)]. The two Drude contributions to the itinerant response are plotted in Fig.~\ref{fig:dispan} as black dashed lines and their plasma frequencies and quasiparticle scattering rate are listed in Table~\ref{table:drudelorentz}. The total plasma frequency of the free electrons $\omega_{\mathrm{pl}}^{\mathrm{tot}}=\sqrt{\omega_{\mathrm{pl,1}}^2+\omega_{\mathrm{pl,2}}^2}$ shows a systematic increase with decreasing atomic number of the intercalant and is significantly larger in $\textrm{CaFe}_2\textrm{As}_2$ than in the other two compounds (see Fig.~\ref{fig:dispan} and Table~\ref{table:drudelorentz}). This trend is also weakly present in first-principles calculations of the plasma frequency~\cite{Dresden_plasma_frequency_review} but is significantly enhanced by renormalization in the real materials.

To futher confirm the values of the total plasma frequency inferred from our Drude-Lorentz fit, we carried out a Kramers-Kronig consistency analysis of ellipsometric data. This analytical technique, utilizing the unique capacity of ellipsometry to independently obtain the real and imaginary part of the dielectric function,~\cite{CharnukhaNatCommun2011} allows one to accurately determine the spectral weight contained in the extrapolation region at lowest frequencies and thus accurately measure the total plasma frequency of itinerant charge carriers. This is made possible by comparing the Kramers-Kronig transformation of the extrapolated real part of the optical conductivity with the independently obtained real part of the dielectric function, as shown in Fig.~\ref{fig:anomalyvis}(e) and the inset therein. The total plasma frequencies for all three compounds obtained with this method are shown in Table~\ref{table:drudelorentz}. It is clear that the agreement between $\omega_{\mathrm{pl}}^{\mathrm{tot}}$ obained using the Drude-Lorentz fit and by means of a Kramers-Kronig consistency check is remarkable. It should be noted that the accuracy of the Kramers-Kronig consistency check in determining the total spectral weight and thus plasma frequency is extremely high (below $0.5\%$ in the present case) and this approach only very weakly depends on the shape of extrapolation.~\cite{CharnukhaNatCommun2011}

The systematic enhancement of the itinerant response with decreasing atomic number of the intercalant, together with the results of the previous sections, shows that a number of electronic propertites are significantly enhanced in $\textrm{CaFe}_2\textrm{As}_2$ compared to its \hbox{Sr-} and Ba-based counterparts.

\section{Conclusions}
In summary, we have reported a systematic comparison of the charge dynamics in three representative parent compounds of 122-type iron-based superconductors ($\textrm{CaFe}_2\textrm{As}_2$, $\textrm{SrFe}_2\textrm{As}_2$, and $\textrm{BaFe}_2\textrm{As}_2$) in a broad spectral range as well as a detailed temperature dependence of their far-infrared conductivity. We identified two spin-density--wave energy gaps in all three compounds, with a significantly stronger coupling in the Ca-based material within both spin-density--wave subsystems as compared to the other two compounds, in which the larger gap decreases systematically with increasing atomic number of the intercalant. Our detailed temperature measurements of the far-infrared conductivity of the three compounds allowed us to accurately track the temperature dependence of both spin-density--wave gaps and reveal an anomaly in $\textrm{CaFe}_2\textrm{As}_2$ in the energy range of the smaller gap at $T^*\approx80\ \textrm{K}$, well below the N\'eel temperature of this compound. The presence of this remnant transition implies that the two spin-density--wave subsystems are very weakly coupled to one another. A comparison to the temperature dependence of the far-infrared conductivity of $\textrm{SrFe}_2\textrm{As}_2$ and $\textrm{BaFe}_2\textrm{As}_2$ shows a clear evolution of this coupling from weak in the Ca-based compound via intermediate in the Sr-based to strong in the Ba-based material. The temperature dependence of the infrared phonons reveals a clear anomaly at $T_{\mathrm{N}}$ in all three compounds, whereby the spin-density--wave--induced modifications of the phonon properties set in exactly at $T_{\mathrm{N}}$ in both \hbox{Ba-} and Sr-based materials but at somewhat higher temperatures in $\textrm{CaFe}_2\textrm{As}_2$, suggesting the early development of spin-density--wave fluctuations and their sizable impact on the lattice. The temperature dependence of the intensity $\Delta\varepsilon_0$ of the Ca-related phonon shows a possible signature of $T^*$. Similarly to the previously reported spin-density--wave--induced suppression of two absorption bands in $\textrm{SrFe}_2\textrm{As}_2$, we find an analogous effect in both \hbox{Ca-} and Ba-based compounds, as well as a systematic evolution of the intensity of the suppression with the atomic number of the intercalant. The investigation of the spectral-weight transfer induced by the spin-density--wave transition reveals that in all three parent compounds energies much larger than the characteristic magnetic energy scales proposed for these materials are involved. This high-energy modification of the spectral weight might originate in a redistribution of the electronic population between several bands at the transition. Our accurate dispersion analysis of the independently obtained real and imaginary parts of the optical conductivity by means of spectroscopic ellipsometry reveals a strongly enhanced plasma frequency in $\textrm{CaFe}_2\textrm{As}_2$, decreasing systematically with increasing atomic number of the intercalating atom. Our results single out $\textrm{CaFe}_2\textrm{As}_2$ in the class of $\textrm{ThCr}_2\textrm{Si}_2$-type iron-based materials by demonstrating the existence of two weakly coupled but extremely metallic electronic subsystems. Since the three investigated materials are driven superconducting upon doping, our results may provide insights into the physics of the related superconductors.

\section{Acknowledgements}

This project was supported by the German Science Foundation under grant BO 3537/1-1 within SPP 1458. We gratefully acknowledge Y.-L. Mathis for support at the infrared beamline of the synchrotron facility ANKA at the Karlsruhe Institute of Technology and V.~Khanna for taking part in some of the measurements.

%\bibliographystyle{C:/PRBbibstyle}
%\bibliography{Z:/Documents/FeAsSC}
%------------- THEBIBLIOGRAPHY ------------------------

%------------- THEBIBLIOGRAPHY ------------------------
\end{document}